\documentclass[preprint2]{aastex}

\newcommand{\rmd}{{\rm d}}

\newcommand{\com}[1] {{\bf #1}}
\newcommand{\etal}{et al.}
\newcommand{\eg}{{\it e.g.}}
\newcommand{\cf}{{\it c.f.}}
\newcommand{\ie}{{\it i.e.}}

\usepackage{amsmath,amssymb}  
\long\def\symbolfootnote[#1]#2{\begingroup 
\def\thefootnote{\fnsymbol{footnote}}\footnote[#1]{#2}\endgroup}

\slugcomment{~}

\shorttitle{Weak lensing in strong lensing regimes}
\shortauthors{R.\ Massey \& D.\ M.\ Goldberg}

\begin{document}

\title{Weak lensing ellipticities in a strong lensing regime}
\author{Richard Massey\altaffilmark{1} \& David M.\ Goldberg\altaffilmark{2}
\altaffiltext{1}{California Institute of Technology, 1200 East California Boulevard, 
Pasadena, CA 91125, U.S.A.}
\altaffiltext{2}{Department of Physics, Drexel University, 3141 Chestnut Street, 
Philadelphia, PA 19104, U.S.A.}
}

\begin{abstract}

It is now routine to measure the weak gravitational lensing shear
signal from the mean ellipticity of distant galaxies. However,
conversion between ellipticity and shear assumes local linearity of
the lensing potential (\ie\ that the spatial derivatives of the shear
are small), and this condition is not satisfied in some of the most
interesting regions of the sky. We extend a derivation of lensing
equations to include higher order terms, and assess the level of
biases introduced by assuming that first-order weak lensing theory
holds in a relatively strong shear regime. We find that, even in a
worst-case scenario, a fully linear analysis is accurate to within
$1\%$ outside $\sim 1.07$ times the Einstein radius of a lens, by
deriving an analytic function that can be used to estimate the
applicability of any first-order analysis.  The effect is too small to
explain the disc\-repancy between weak- and strong-lensing estimates of
the mass of the bullet cluster, and should not impact cluster surveys
for the forseeable future. In fact, it means that arclets can be used
to measure shears closer to a cluster core than has been generally
appreciated.  However, at the level of accuracy demanded by future
lensing surveys, this bias ought to be considered in measure\-ments of
the inner slope of cluster mass distributions and the small-scale end
of the mass power spectrum. Both of these are central in determining
the relationship between baryonic and dark matter.

\end{abstract}

\keywords{gravitational lensing}

\section{Introduction}

Gravitational lensing is the deflection of light rays from a background light
source by an intervening gravitational field \citep{melrev,refrev}. It is one of
the most promising probes of the distribution of dark matter, and hence the
effects of dark energy. Along lines of sight where the deflection is sufficient, ``strong
lensing'' visibly distorts (and often multiply images) the shapes of individual background 
galaxies. However, only ``weak lensing'' is produced along most lines of sight, even those passing through the outskirts of
galaxy clusters. This weaker but ubiquitous signal has to 
be collected statistically. To first order in a Taylor series, it is obtained
from the mean ellipticity of an otherwise uncorrelated set of galaxies
\citep{bs}.

Weak
lensing measurements have now been well used to map the distribution of mass
\citep{bullet,CFHTmap,cosmos_map} and characterize its large-scale statistical
properties \citep{cosmos_cs,benjamin,kitching3d}. However, it is often the most
massive structures that are of particular interest in the maps
\citep[\eg][]{witmanclusters,gabodsclusters,satoshiclusters}, and that dominate
the contribution to the power spectrum on small scales \citep[\eg][]{smithps}.
Near such regions, the first-order assumptions implicit in a weak lensing analysis no longer
necessarily hold. In this paper, we expand the Taylor series of the weak lensing
equation to include the next-highest terms, and investigate the level of bias in
shear measurements that rely upon simple measurements of ellipticity. 

We derive the lensing equations in \S\ref{sec:transforms}.
We check our results using raytraced simulations in \S\ref{sec:raytrace}, 
and we discuss their implications in \S\ref{sec:conc}.

\section{Lensing Transformations}
\label{sec:transforms}

\subsection{The Usual First-Order Treatment}
\label{sec:firstorder}

A general gravitational lens deflects a light from a position $x^\prime$ in a background 
(source) image to a position $x$ in the
observed (lens) plane, such that
\begin{equation}
\vec{x^\prime}=\vec{x}-\vec{\alpha}(\vec{x}) ~,
\label{eqn:lensequation}
\end{equation}
with a deflection angle predicted by General Relativity in the weak field limit of
\begin{equation}
\vec{\alpha}(\vec{x})=\vec{\nabla}\Psi(\vec{x}) ~,
\label{eqn:grprediction}
\end{equation}
and where $\Psi(\vec{x})$ is the Newtonian potential of the lens, $\Phi(\vec{x},z)$, 
projected onto the 
plane of the sky.

Crucially, the gravitational field and the deflection angle vary across the sky.
Assuming (the local linearity condition) that the change is linear on scales 
the size of a galaxy, it can be described to first order by a coordinate transformation
\begin{equation}
x^\prime_i = x_i-\left[\frac{\partial \Psi}{\partial x_i} \right] -
\frac{\partial}{\partial x_j}\left[\frac{\partial \Psi}{\partial x_i}\right]\Delta x_j +... 
~.
\label{eqn:lineartrans}
\end{equation} The first derivative term represents an unmeasurable
centroid shift. Placing the origin of the coordinate system at the
galaxy's observed center of light, we are left with 
\begin{equation} 
x^\prime_i = \mathcal{A}_{ij} x_j + ...~,
\end{equation}
where the Jacobian of the transformation is
\begin{eqnarray}
\mathcal{A}_{ij}
& = & \delta_{ij} -
\frac{\partial^2\Psi}{\partial x_i\partial x_j} \\
\mathcal{A} & \equiv & \left(
\begin{array}{cc}
  1-\kappa-\gamma_1 & -\gamma_2 \\
  -\gamma_2 & 1-\kappa+\gamma_1 \\
\end{array}
\right) ~.
\label{eqn:Amatrix}
\end{eqnarray}
We have introduced the usual notation of convergence $\kappa(\vec{x})=\vec{\nabla}^2\Psi(\vec{x})/2$,
which is proportional to the distribution of mass projected along a line of sight,
and two components of 
shear $\gamma_i(\vec{x})$. The inverse mapping is simply
\begin{equation}
x_i=(\mathcal{A})^{-1}_{ij}x^\prime_j+...~.
\label{eqn:linearinvtrans}
\end{equation}

It is always possible to adopt an arbitrary choice of rotation for the coordinate system 
such that $\gamma_2=0$ (so $\mathcal{A}$ is diagonal), 
and invoke parity symmetry to consider only that the potential increases to the right (hence
$\gamma_1<0$). We also work only in the ``positive parity'' lensing regime
(outside the critical curve), where $\mathrm{det}\mathcal{A}>0$.
Our analysis is equally valid inside the critical curve, although breaks down
if a part of the image crosses the critical curve \citep[\cf][]{bananas}.

The shape of a galaxy image $I(x^\prime)$ can be quantified via its intrinsic ellipticity
\begin{equation}
\left\{\chi_1^\mathrm{int},\chi_2^\mathrm{int} \right\} ~\equiv~
\left\{\frac{Q_{11}^\mathrm{int}-Q_{22}^\mathrm{int}}{Q_{11}^\mathrm{int}+Q_{22}^\mathrm
{int}},\frac{2Q_{12}^\mathrm{int}}{Q_{11}^\mathrm{int}+Q_{22}^\mathrm{int}}\right\} ~,
\end{equation}
where its quadrupole moments are
\begin{eqnarray}
Q_{ij}^\mathrm{int}\equiv\frac{\int I(\vec{x^\prime})~x^\prime_i~x^\prime_j~\rmd^2\vec{x^
\prime}}{\int I(\vec{x^\prime})~\rmd^2\vec{x^\prime}} ~.
\label{eqn:quadrupoledefn}
\end{eqnarray}

Under the (locally linear) lensing transformation~\eqref{eqn:linearinvtrans}, the galaxy's 
observed ellipticity becomes
\begin{equation}
\chi_i^\mathrm{obs}=\chi_i^\mathrm{int}+2\gamma_i-\chi_i^\mathrm{int}(\chi_j^\mathrm{int}
\gamma_j)~,
\end{equation}
to first order in $\gamma$ \citep{seitzschneider}. 
Averaging over a population of galaxies with
uncorrelated intrinsic shapes $\langle\chi^\mathrm{int}\rangle=0$, 
an estimator 
$\tilde{\gamma}$ can then  recover the gravitational shear signal
\begin{equation}
\left\langle\tilde{\gamma_i}\right\rangle\equiv
\frac{\left\langle\chi_i^\mathrm{obs}\right\rangle}{2-\left\langle(\chi_i^\mathrm{int})^2
\right\rangle} 
=\left\langle\gamma_i\right\rangle ~.
\label{eqn:shearestimator}
\end{equation}
The variance in the denominator can be closely approximated by the observed value. 
It is typically of order 0.4 \citep[\eg][]{alexie}.

For practical purposes, a weight function $W(\vec{x})$ with finite support is
also usually applied to the integrals in equation~\eqref{eqn:quadrupoledefn}.
This complicates the shear estimator: the shear
polarizability tensor $P^\gamma$ in \citet{ksb}, which generalizes the denominator
of equation~\eqref{eqn:shearestimator}, involves derivatives of
$W(\vec{x})$. However, $P^\gamma$ is typically fitted from a large ensemble of
galaxy shapes to reduce noise, and almost all of those galaxies will be on
lines of sight unaffected by higher order lensing terms. We therefore ignore the effect 
here\footnote{As pointed out during the derivation of ``reduced shear'' by
\citet{bs}, a galaxy's flux $I(\vec{x'})$ could be replaced
in eq.~\eqref{eqn:quadrupoledefn} and throughout by a monotonic function of
intensity $f(I(\vec{x'}))$, without any change in the formalism. This approximates
a useful weighting scheme.}. 

\subsection{Higher order terms}
\label{sec:higherorder}

\noindent Continuing the Taylor series in equation~\eqref{eqn:lineartrans}, we can write
\citep[\cf][]{flexion1}
\begin{eqnarray}
x^\prime_i & = & \mathcal{A}_{ij} x_j 
               - \frac{1}{2}\frac{\partial^3\Psi}{\partial x_i\partial x_j\partial x_k}
x_jx_k \nonumber  \\
      &~& ~~~  - \frac{1}{6}\frac{\partial^4\Psi}{\partial x_i\partial x_j\partial x_k
\partial x_l}x_jx_kx_l +... 
\label{eqn:highertrans}
\end{eqnarray}
Repeatedly substituting the simple form
\begin{equation}
x_i = (\mathcal{A})^{-1}_{ij} \left( x^\prime_j
      +\frac{1}{2}\Psi,_{jkl}x_kx_l +\frac{1}{6}\Psi,_{jklm}x_kx_lx_m\right)
\label{eqn:higherinvtrans}
\end{equation}
into itself then yields the inverse mapping
\begin{eqnarray}
x_i & = & (\mathcal{A})^{-1}_{ij} x_j^\prime \\
    & + &    ^1/_2 (\mathcal{A})^{-1}_{ij} (\mathcal{A})^{-1}_{kp} (\mathcal{A})^{-1}_{lq} 
\Psi,_{jkl} x^\prime_p x^\prime_q \nonumber \\
    & + &    ^1/_6 (\mathcal{A})^{-1}_{ij} (\mathcal{A})^{-1}_{kp} (\mathcal{A})^{-1}_{lq} 
(\mathcal{A})^{-1}_{mr} \Psi,_{jklm} x^\prime_p x^\prime_q x^\prime_r \nonumber \\
    & + &    ^1/_2 (\mathcal{A})^{-1}_{ij} (\mathcal{A})^{-1}_{kp} (\mathcal{A})^{-1}_{lm} 
(\mathcal{A})^{-1}_{nq} (\mathcal{A})^{-1}_{sr} \nonumber\\
    & ~ &             ~~~~~~~~~~~~~~~~~~~      \Psi,_{jkl} \Psi,_{mns} x^\prime_p 
x^\prime_q x^\prime_r ~~ + ~~ ... \nonumber
\end{eqnarray}
The various terms are listed in order of decreasing importance. 
Third derivatives of $\Psi$ are related to the {\em flexion} signal \citep
{flexion2,flexion3}.
This is small for realistic potentials; higher derivatives of $\Psi$ will be smaller still.
Note that this relation still holds locally even if there are
multiple images, but that there will be different values of $\mathcal{A}$ at each image.

To complicate matters, this mapping now shifts the galaxy's center of light.
If the background image were correctly centroided  (\ie\  
$\langle x^\prime\rangle=0$), the observed centroid would be
\begin{equation}
\langle x_i\rangle \approx \frac{1}{2}(\mathcal{A})^{-1}_{ij} (\mathcal{A})^{-1}_{km}
(\mathcal{A})^{-1}_{ln} \Psi,_{jkl} Q_{mn}^\mathrm{int}~,
\end{equation} 
plus smaller contributions coming from the galaxy's intrinsic
octopole moment.  In a coordinate system centered on the observed
image, the mapping (for a fully general potential) is therefore (\cf\ eq. 
\ref{eqn:linearinvtrans})
\begin{eqnarray} \label{eq:finalcoord}
x_i & = & (\mathcal{A})^{-1}_{ij} x_j^\prime \\
 & + & ^1/_2 (\mathcal{A})^{-1}_{ij}
(\mathcal{A})^{-1}_{kp} (\mathcal{A})^{-1}_{lq} \Psi,_{jkl}
\big(x^\prime_p x^\prime_q-Q_{pq}^\mathrm{int}\big) \nonumber \\ & + &
^1/_6 (\mathcal{A})^{-1}_{ij} (\mathcal{A})^{-1}_{kp}
(\mathcal{A})^{-1}_{lq} (\mathcal{A})^{-1}_{mr} \Psi,_{jklm}
x^\prime_p x^\prime_q x^\prime_r \nonumber \\ & + & ^1/_2
(\mathcal{A})^{-1}_{ij} (\mathcal{A})^{-1}_{kp}
(\mathcal{A})^{-1}_{lm} (\mathcal{A})^{-1}_{nq}
(\mathcal{A})^{-1}_{sr} \nonumber\\ 
 & ~ &             ~~~~~~~~~~~~~~~~~~~      \Psi,_{jkl} \Psi,_{mns} x^\prime_p 
x^\prime_q x^\prime_r ~~ + ~~ ... \nonumber
\end{eqnarray} 
In practice, a galaxy's intrinsic quadrupole moments cannot be observed. We expand them as a 
function of the
galaxy's observed shape using equation~\eqref{eqn:highertrans}.  However, several 
non-negligible coefficients produce an unwieldly general
expression.

To make the equations more tractable, we now fix various properties of the lens and the source galaxy. 
We first set to zero all derivatives of $\Psi$ that are ``odd'' at $90^\circ$
($\Psi,_{112}$, $\Psi,_{222}$, $\Psi,_{1112}$ and $\Psi,_{1222}$). For a circular (or nearly 
circular) potential that has
been rotated so that $\Psi,_{12}=0$, this assumption will be (nearly) accurate. 
It is also explicitly true at the major and minor axes of an elliptical potential. 

Since we are in a fairly strong lensing regime, it is not unreasonable to assume
that $\gamma\gg\chi^\mathrm{int}$, so the galaxy can be considered intrinsically circular. 
It still has a size $R^2\equiv 2Q_{11}^\mathrm{int}=2Q_{22}^\mathrm{int}$ and concentration 
index 
\begin{equation}
c\equiv\frac{\int I(\vec{x})~|\vec{x}|^4~\rmd^2\vec{x}}{(R^2)^2\int I(\vec{x})~\rmd^2\vec{x}} ~,
\end{equation} 
which would be 2 for a Gaussian, 10/3 for an exponential, and higher still for a de 
Vaucouleurs profile.
The observed ellipticity becomes
\begin{eqnarray}
\label{eqn:nonlinearchi}
\chi_1^\mathrm{obs}= \chi_1^\mathrm{lin} - \frac{a^2 d^2 ~~ R^2}{4(a^2+d^2)^2} 
\Big[\big\{ a^2 \Psi,_{111} + d^2 \Psi,_{122} \big\}^2~~~~ \\
-c\big\{
15   a^4                            \Psi,_{111}^2             
-(12 a^2 d^2              
+  4 a   d^3 
-  3 d^4      )\Psi,_{122}^2         ~~~~~~~\nonumber \\                  
- 2a^2d(2a-3d) \Psi,_{111}   \Psi,_{122} ~~~~~~~ \nonumber \\
+  4 a^3                            \Psi,_{1111}        
-  4 a d (a-d) \Psi,_{1122}
-  4 d^3                            \Psi,_{2222}           
\big\}\Big] ~.~   \nonumber
\end{eqnarray}
where $a\equiv(\mathcal{A}^{-1})_{11}=(1-\Psi,_{11})^{-1}$ and $d\equiv(\mathcal{A}^{-1})_
{22}=(1-\Psi,_{22})^{-1}$ are  unitless.
For a Singular Isothermal Sphere (SIS) lens, $\Psi(\vec{x})=\theta_E|\vec{x}|$, 
\begin{equation}
\chi_1^\mathrm{obs}= \chi_1^\mathrm{lin} -
\frac{cR^2}{\theta_E^2}
\frac{\left[12r^3-(7-\frac{1}{c})r^2-12r-7\right]}{4\left(r-1\right)^{4}(r^2+\left(r-1\right)^2)}
\end{equation}
where $r\equiv|\vec{x}|/\theta_E$. The deviation from an ellipticity assuming local linearity, 
$\chi^\mathrm{lin}$, tends as $R^2/\theta_E^2$.

\section{Verification through raytracing}
\label{sec:raytrace}

\begin{figure}[tb]
\epsscale{1}
\plotone{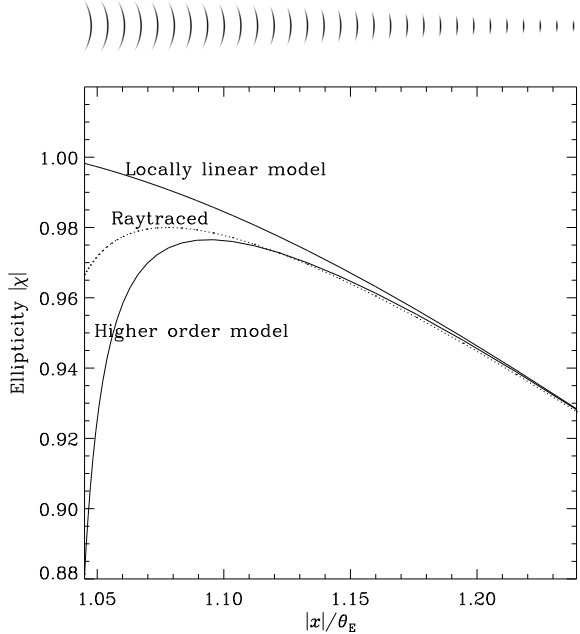}
\epsscale{1}
\caption{
{\it Upper images}: The observed shape of an intrinsically circular galaxy with a Gaussian 
radial 
profile and size $\sigma=0.01\theta_E$, at various positions behind a singular isothermal 
sphere lens. 
The images are presented with a logarithmic color stretch.
{\it Main panel}: 
The solid lines show the object's ellipticity predicted by the usual linear model and our 
higher order model. 
The dotted line shows measured values from a fully raytraced simulation.}
\label{fig:chi_vs_re}
\end{figure}

We have developed a simple raytracing routine to deflect rays via equation~\eqref
{eqn:grprediction}, deforming
the intrinsic shapes of source galaxies into arcs. The upper panel of figure~\ref
{fig:chi_vs_re} demonstrates
the effect of a singular isothermal sphere lens with Einstein radius $\theta_E$ on an 
intrinsically circular
Gaussian source with $\sigma=0.01\theta_E$. Note that this is a worst-case scenario in 
several respects, with
more concentrated or smaller galaxies being less affected.  If the lens were Abell 1689, 
this would correspond
to a $z=1$ galaxy of FWHM $\sim 1\arcsec$ \citep{clowea1689}, which is amongst the largest 1
\% of
\citet{alexie}'s catalog at magnitude $i^\prime=25$.

The main panel of figure~\ref{fig:chi_vs_re} shows the measured ellipticity of the raytraced 
images, and the
prediction of linear and higher order models. These converge away from the lens; the slight
difference between them and the raytraced version is an effect of image pixellization. 
converging slowly.
Near the lens, our nonlinear model~\eqref{eqn:nonlinearchi} again presents a worst
case of the deviation from a linear prediction. It differs from the raytraced
measurements due to even higher order terms in the coordinate transformation.

\section{Discussion}
\label{sec:conc}

We have derived the next-highest terms in the coordinate transformation relevant for weak 
gravitational lensing,
by dropping the assumption of ``local linearity'', which acts as a
useful constraint on the applicability of the linear approximation.
The resulting equations are not elegant, but can be
simplified by making several reasonable assumptions about the galaxy's intrinsic shape and 
the lens profile.
As expected, the perturbations from linear lensing theory are greatest
for large galaxies; they increase as the size of the galaxy
squared. Like with gravitational flexion, this is simply due to the
accumulating change in shear signal across the width of an image.

A linear lensing analysis systematically overestimates the shear signal near the core of 
galaxy clusters.
However, even in the worst case scenario, it is acceptable surprisingly far into the non-
linear regime. Assuming
a value of 1.6 for the denominator of equation~\eqref{eqn:shearestimator}, it is within 1\% 
of the true shear
outside $\sim 1.07\theta_E$\com, where $\gamma\simeq 0.47$, and the
reduced shear, $g\simeq 0.93$.
Compared to other potential errors, we therefore conclude that this will be of only minor 
concern for measurements of
the mass of individual (or even stacked) clusters in immediately forthcoming surveys. 
For example, the effect is in the right direction but an order of magnitude too small to 
explain the discrepancy between
measurements of the mass in the bullet cluster \citep{bullet} via strong and weak lensing.
However, the effect ought to be considered by programs measuring the inner slopes of
cluster mass distributions or the mass power spectrum on small scales. The effect can become 
relevant at about
the level of statistical accuracy proposed for next-generation surveys. 

We have not investigated the correction for a point spread function or the use of a weight 
function while
measuring galaxy shapes. 
A full analysis of these would be interesting in future work.

\acknowledgments

The authors thank Douglas Clowe, Yannick Mellier and Barnaby Rowe for useful discussions. 
This work was supported by NASA grant ATP04-0000-0067 and DoE grant FG02-04ER41316.




\end{document}